\documentclass[sigconf]{acmart}
\usepackage{tabularx}
\newcolumntype{Y}{>{\centering\arraybackslash}X}

\usepackage{algorithm}
\usepackage{algpseudocode}
\usepackage{multirow}
\usepackage{subcaption}
\usepackage{cleveref}

\usepackage{array}
\usepackage{makecell}

\algdef{S}[FOR]{ForEach}[1]{\algorithmicforeach\ #1\ \algorithmicdo}

\copyrightyear{2024} 
\acmYear{2024} 
\setcopyright{acmcopyright}
\acmConference[CIKM '24]{Proceedings of the 31st ACM International Conference on Information and Knowledge Management}{May 7--11, 2024}{Atlanta, GA, USA}
\acmBooktitle{Proceedings of the 31st ACM International Conference on Information and Knowledge Management (CIKM '24), May 7--11, 2024, Atlanta, GA, USA}
\acmPrice{15.00}
\acmDOI{10.1145/3511808.3557065}
\acmISBN{978-1-4503-9236-5/24/10}
\begin{document}

\title{Dynamic User Interest Augmentation via Stream Clustering and Memory Networks in Large-Scale Recommender Systems} 

\author{Peng Liu}
\email{liupengvswww@gmail.com}
\orcid{0009-0000-7271-4721}
\affiliation{%
  \institution{Tencent Inc.}
  \city{Beijing}
  \country{China}
}

\author{Nian Wang}
\email{noreenwang@tencent.com}
\affiliation{%
  \institution{Tencent Inc.}
  \city{Beijing}
  \country{China}
}

\author{Cong Xu}
\email{congcxu@tencent.com}
\affiliation{%
  \institution{Tencent Inc.}
  \city{Beijing}
  \country{China}
}

\author{Ming Zhao}
\email{marcozhao@tencent.com}
\affiliation{%
  \institution{Tencent Inc.}
  \city{Beijing}
  \country{China}
}

\author{Bin Wang}
\email{hillmwang@tencent.com}
\affiliation{%
  \institution{Tencent Inc.}
  \city{Beijing}
  \country{China}
}

\author{Yi Ren}
\email{henrybjren@tencent.com}
\affiliation{%
  \institution{Tencent Inc.}
  \city{Beijing}
  \country{China}
}

\renewcommand{\shortauthors}{Peng Liu et al.}

\begin{abstract}
Recommender System (RS) provides personalized recommendation service based on user interest. 
However, lots of users' interests are sparse due to lacking consumption behaviors, 
making it challenging to provide accurate recommendations for them, which is widespread in large-scale RSs. 
In particular, efficiently solving this problem in the ranking stage of RS is an even greater challenge, which requires an end-to-end and real-time approach.

To solve this problem, we propose an innovative method called Dynamic User Interest Augmentation (DUIA). 
DUIA enhances user interest including user profile and user history behavior sequences by generating enhancement vectors and 
personalized enhancement vectors through dynamic stream clustering of similar users and relevant items from multiple perspectives. 
To realize stream clustering, we specially design an algorithm called Gradient-based Hierarchical Clustering Algorithm (GHCA) for DUIA, 
which performs clustering via gradient descent and stores the cluster centers in memory networks. 
Extensive offline and online experiments demonstrate that DUIA not only significantly improves model performance for users with sparse interests 
but also delivers notable gains for other users. As an end-to-end method, DUIA can be easily integrated with existing models. 
Furthermore, DUIA is also used for long-tail items and cold-start problem, which also yields excellent improvements. 
Since 2022, DUIA has been successfully deployed in multiple industrial RSs in Tencent and was made public in May 2024. 
Moreover, the thoughts behind DUIA, dynamic stream clustering and similarity-based enhancement, 
have inspired relevant works and have also been applied in other stages of RS.
\end{abstract}



\maketitle

%%%%%%%%%%%%%%%%%%%%%%%%%%%%%%%%%%%%%%%%%%%%%%%%%%%%%%%%%%%%%%%%%%%%%%%%%%%%%%%%%%%%%%
%%%%%%%%%%%%%%%%%%%%%%%%%%%%%%%%       Intro      %%%%%%%%%%%%%%%%%%%%%%%%%%%%%%%%%%%%
%%%%%%%%%%%%%%%%%%%%%%%%%%%%%%%%%%%%%%%%%%%%%%%%%%%%%%%%%%%%%%%%%%%%%%%%%%%%%%%%%%%%%%

\section{Introduction}
\label{sec:intro}
\begin{figure}[hbtp!]
  \centering
  \includegraphics[width=0.85\linewidth]{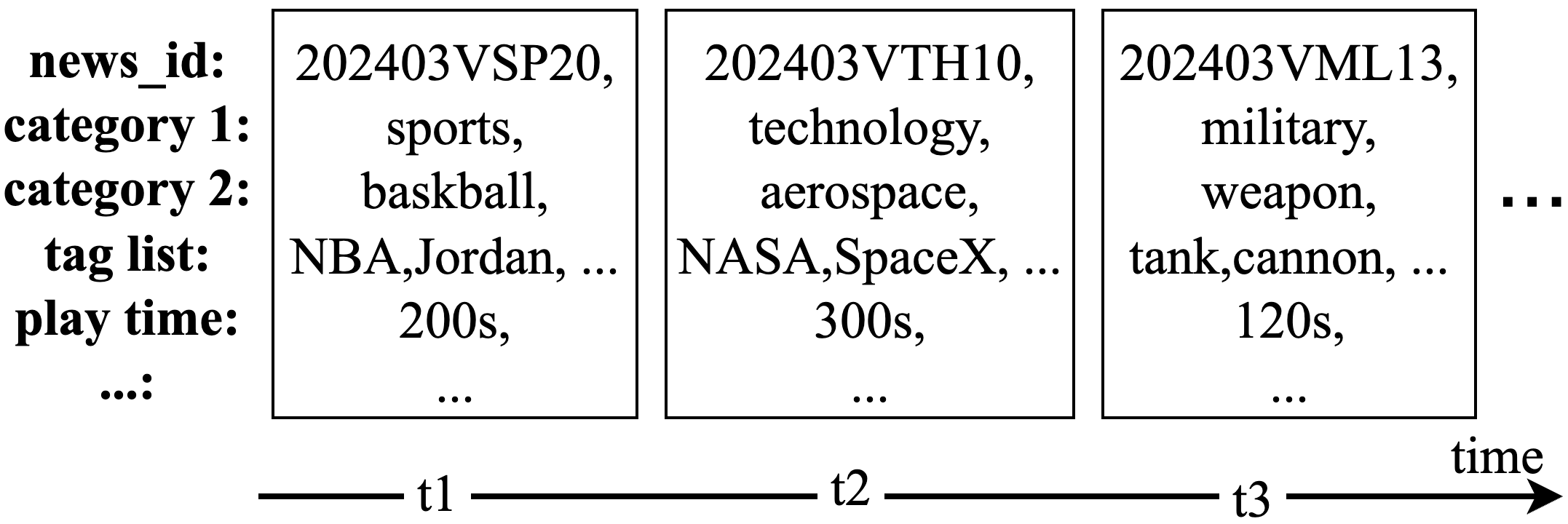}
  \caption{An example of user valid consumption sequence.}\label{fig:f2}
  \vspace{-0.1in}
\end{figure}

% [recsys is popular] 
Recommender System (RS) \cite{ref1, ref2} that provides recommendation service based on user interest is widely used in various platforms 
such as short video platforms \cite{ref3, ref7, ref14}, video platforms \cite{ref4, ref5}, E-commerce platforms \cite{ref6, ref8, ref9, ref10, ref11} 
and social networks \cite{ref12, ref13}, serving billions of users. In the ranking stage of RS, 
a Multi-Task Learning (MTL) model \cite{ref4, ref8, ref16, ref17, ref18, ref19, ref20, ref21} is used for predicting the scores of 
various user behaviors using elaborate model structure and lots of features, which is critical for personalized RSs.
User interest includes user profile and user history behavior sequences, as shown in Table \ref{table:user_profile} and Figure \ref{fig:f2}, 
which determines the upper bound of the MTL model's performance. 

However, a large amount of users only have sparse interest (low-active users) due to lacking consumption behaviors. 
For example, users with sparse interest account for more than 30\% of the daily active users in our RS, 
and their average number of consumption behaviors is only half that of high-active users. 
As a result, this makes it challenging for MTL model to accurately estimate scores for these users, leading to poor recommendation. 
It is widespread in large-scale RSs. Improving recommendation accuracy for low-active users is very important, 
but also particularly difficult.

Although there have been considerable works on MTL \cite{ref4, ref8, ref16, ref17, ref18, ref19, ref20, ref21}, 
studies focusing on users with sparse interest are not many and they have not achieved remarkable improvements. 
To improve model performance for users with sparse interest, we propose a novel method called Dynamic User Interest Augmentation (DUIA). 
DUIA enhances user interest, including user profile and historical behavior sequences, by enhancement vectors and personalized enhancement vectors 
generated with the help of similar users and relevant items through dynamic stream clustering and memory networks. 
To realize that efficiently, DUIA does not directly adopt existing clustering algorithms, such as K-means, Divisive \cite{ref37} or VQ-VAE \cite{ref39}, 
as these methods are either unsuitable for streaming clustering or are specially designed for the requirements of Computer Vision (CV), 
and therefore do not fully meet the requirements of DUIA. 
Instead, we design the Gradient-based Hierarchical Clustering Algorithm (GHCA) for DUIA, which performs clustering via gradient descent and 
stores the cluster centers in memory networks, resulting in higher efficiency and better performance. 
DUIA significantly improves model performance for users with sparse interest by producing pivotal interest information. 
Moreover, DUIA also significantly improves model performance for other users. 
Similarly, we further apply DUIA to long-tail items and cold-start problem, achieving excellent improvements as well.

In this paper, our contributions can be summarized as follows:
\begin{itemize}
  \item  We first point out that a large number of users only have sparse interest, which is common and 
  results in poor recommendation. Moreover, a similar problem also exists for long-tail items due to highly skewed distribution.
    
  \item We propose a novel method called DUIA for MTL, generating user interest enhancement vectors through dynamic stream clustering and 
  similarity-based augmentation. We also specially design a stream clustering algorithm called GHCA for DUIA. 

  \item Extensive experiments show that DUIA remarkably outperforms existing methods. 
  For low-active users, DUIA improves User Valid Consumption (UVC) by 9.68\% and User Duration Time (UDT) by 5.74\%. 
  For high-active users, DUIA increases UVC by 5.60\% and UDT by 3.88\%. 
  Moreover, DUIA substantially improves performance on long-tail items and mitigates the cold-start problem.

  \item DUIA has been deployed in multiple large-scale RSs since 2022. Its core ideas—dynamic stream clustering and similarity-based enhancement—have also 
  inspired similar works and been used in candidate generation, pre-ranking, and context-aware DNN.

\end{itemize}
\begin{table}[hbtp!]
  \caption{An example of user profile.}
  \label{table:user_profile}
  \renewcommand\arraystretch{1.2}{
  \begin{minipage}{\columnwidth}
  \begin{center}
  \resizebox{\linewidth}{!}{
  \begin{tabular}{ c|c}
  \toprule
  \small{Basic Attribute} & \thead{ user id, gender, hometown, phone brand, etc.} \\ \hline
  \small{User Preference on Category 1} & \thead{sports:95, technology:90, military:65, etc.} \\ \hline
  \small{User Preference on Category 2} & \thead{baskball:90, football:82, weapon:60, etc.} \\ \hline
  \small{User Preference on Tag} & \thead{NBA:92, CBA:85, AIGC:80, tank:80,\\ space shuttle:70, cannon:50, warplane:30, etc.} \\ \hline
  \small{Others Features} & ... \\
  \bottomrule
  \end{tabular}
  }
  \end{center}
  \end{minipage}
  }
  \vspace{-0.1in}
\end{table}

%%%%%%%%%%%%%%%%%%%%%%%%%%%%%%%%%%%%%%%%%%%%%%%%%%%%%%%%%%%%%%%%%%%%%%%%%%%%%%%%%%%%%%
%%%%%%%%%%%%%%%%%%%%%%%%%%%%%%%%   Related Work   %%%%%%%%%%%%%%%%%%%%%%%%%%%%%%%%%%%%
%%%%%%%%%%%%%%%%%%%%%%%%%%%%%%%%%%%%%%%%%%%%%%%%%%%%%%%%%%%%%%%%%%%%%%%%%%%%%%%%%%%%%%

\section{Related Work}
\label{related_work}
% to check and polish
MTL \cite{ref4, ref8, ref16, ref17, ref18, ref19, ref20, ref21} is responsible for predicting the scores of various user behaviors, 
which is critical for RS. In large-scale RSs such as YouTube, Shorts, Kwai, TikTok, 
a great number of users merely have sparse interest because of lacking consumption behaviors including valid click, watching, like, collect, share etc. 
For example, among the daily active users in our RS, low-activity users account for more than 30\%. 
The features of user interest decide the upper bound of models' performance. 
Therefore, due to lacking key information of user interest including user profile and user history behavior sequences, 
the MTL model performance for low-active users is poor. 

Enhancing model performance for low-active users is necessary but also quite challenging. 
To date, the research focusing specifically on low-active users is not much. 
\cite{ref22, ref23, ref24} attempt to learn initial ID embedding for new items or users based on meta-learning. 
In addition, \cite{ref25, ref26} aim to obtain initial Identifier (ID) embeddings for new users or items via 
Conditional Variational Auto-Encoders (CVAE) \cite{ref25}. 
These methods are primarily designed to address the cold-start problem for new users or items, 
and are not suitable for low-active users, who have some consumption behaviors but far fewer than high-active users. 
\cite{ref28} improves representation learning for sparse data using data augmentation and a Self-Supervised Learning (SSL) framework, 
which can be used to enhance user interest for low-active users. However, in the context of RSs, 
data augmentation in Contrastive Learning (CL) \cite{ref28} can easily generate meaningless data, thus limiting its value to the model. 
\cite{ref29} proposes a dual transfer learning framework that transfers knowledge from head items to tail items or from high-active users to low-active users. 
However, transfer learning is highly sensitive to differences between source and target data distributions, 
and may even lead to negative transfer \cite{ref30}. 
Furthermore, \cite{ref3} uses personalized prior information as input and dynamically scales the bottom-level embedding 
and top-level Deep Neural Network (DNN) hidden units, which can improve model performance for low-active users 
from the perspectives of model structure and embedding. 
Based on similar idea, \cite{ref31} farther optimizes MTL model's performance for low-active users. 
However, due to the lack of key interest information for low-active users, the improvements achieved by these methods are not significant.
In summary, existing works have not achieved remarkably improvements for low-active users.

%%%%%%%%%%%%%%%%%%%%%%%%%%%%%%%%%%%%%%%%%%%%%%%%%%%%%%%%%%%%%%%%%%%%%%%%%%%%%%%%%%%%%%
%%%%%%%%%%%%%%%%%%%%%%%%%%%%%%  Problem Definition  %%%%%%%%%%%%%%%%%%%%%%%%%%%%%%%%%%
%%%%%%%%%%%%%%%%%%%%%%%%%%%%%%%%%%%%%%%%%%%%%%%%%%%%%%%%%%%%%%%%%%%%%%%%%%%%%%%%%%%%%%
\section{User Interest and Data Analysis}
\label{sec:data_analysis}
\begin{table}[hbtp!]
  \caption{The comparative values of the average number of user profile features and the average number of consumption behaviors between 
   low-active users and high-active users.}
  \label{table:user_feats}
  \footnotesize
  \renewcommand\arraystretch{1.3}{
  \begin{minipage}{\columnwidth}
  \begin{center}
  \resizebox{\linewidth}{!}{
  \begin{tabular}{ c|c|c }
  \toprule
  \textbf{ User Type} & \textbf{User Profile} & \textbf{Consumption Behaviors} \\ \hline
  \textbf{High-active Users} & 1.0 & 1.0 \\
  \textbf{Low-active Users} & 0.54 & 0.41 \\
  \bottomrule
  \end{tabular}
  }
  \end{center}
  \end{minipage}
  }
\end{table}

User interest is composed of both user profile and user historical behavior sequences. 
User profile includes basic attributes such as user ID, age, gender and hometown, 
as well as preference features inferred from the user's various behaviors over a period of one or several months, 
as shown in Table \ref{table:user_profile}.
User historical behavior sequences include several types, such as valid consumption sequences and interaction sequences. 
Taking valid consumption sequence as an example, as shown in Figure \ref{fig:f2}. 
A valid consumption is defined as watching a video for more than 10 seconds. 
\begin{figure*}[hbtp!]
    \centering
    \includegraphics[width=1.0\linewidth]{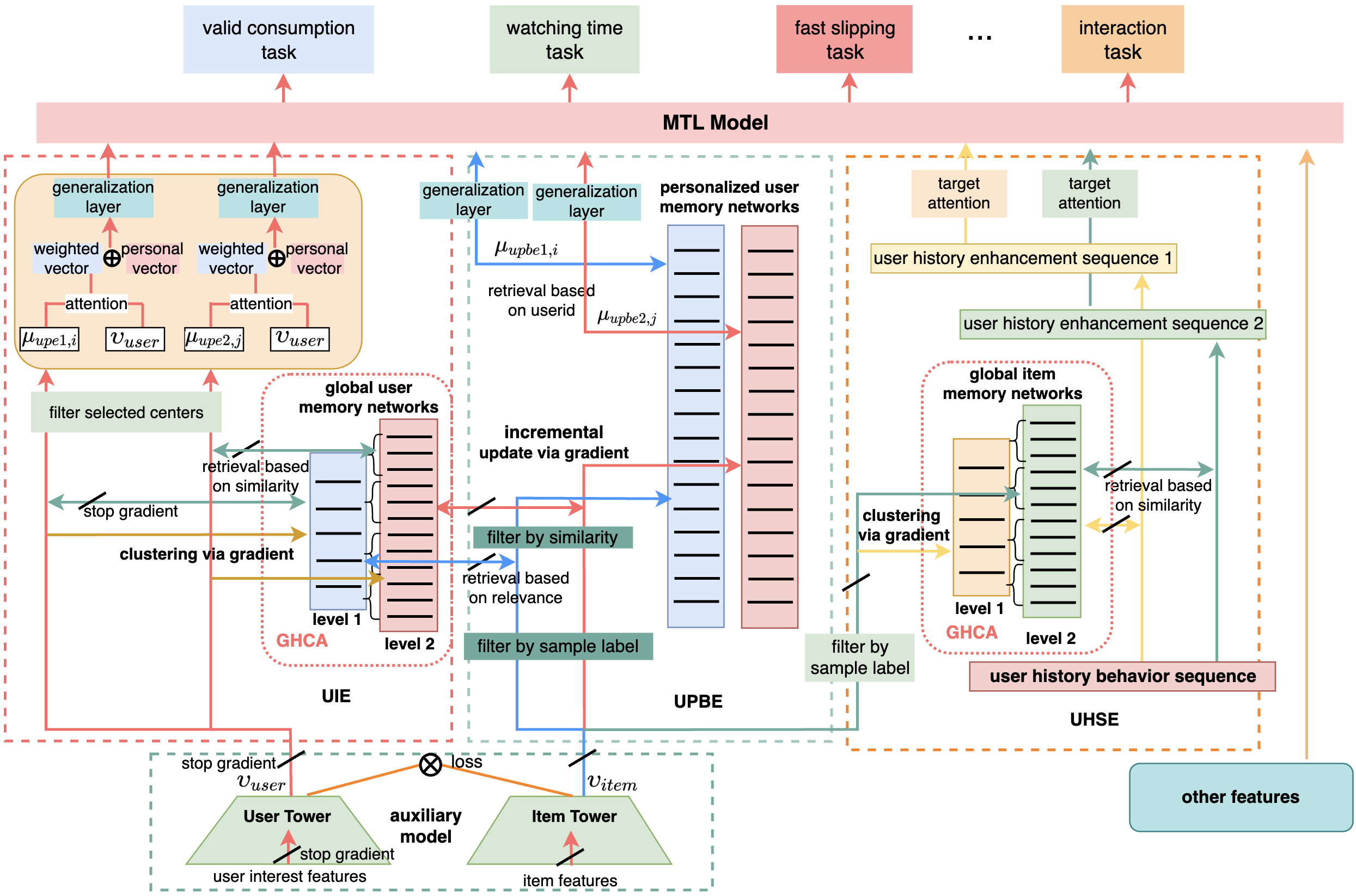}
    \caption{
      % update
      System overview of our ranking model with DUIA, including UIE, UPBE and UHSE. 
      To generate fixed-length $\upsilon_{user}$ and $\upsilon_{item}$, an auxiliary user-item model is defined. Memory networks are used for storage, 
      retrieval and updating of information. All the vectors outputted by these enhancement components are used as new features of 
      ranking model.
    }\label{fig:f3}
    % \vspace{-0.1in}
\end{figure*}

As mentioned before, plenty of users only have spare interest. 
For example, within our RS's daily active user, around 32\% are low-active users, and the other 68\% are high-active users. 
Although the definition of low-active users may differ across different recommendation scenarios, 
the average number of low-active users' interest features is significantly lower than that of high-active users.
In our RS, the average number of low-active users' profile features is only 54\% of that of high-active users and 
the average number of low-active users' consumption behaviors is merely 41\% of that of high-active users, as shown in Table \ref{table:user_feats}. 
Due to low-active users severely lacking interest information, in our RS, 
the test Area Under Curve (AUC) \cite{ref32} of low-active users' valid consumption is approximately 2\% lower than that of high-active users. 
Similarly, highly skewed item distribution is common in RSs. 
In our RS, the top 10\% of items account for over 90\% of total exposures in a single day, which significantly impairs model performance on tail items.

%%%%%%%%%%%%%%%%%%%%%%%%%%%%%%%%%%%%%%%%%%%%%%%%%%%%%%%%%%%%%%%%%%%%%%%%%%%%%%%%%%%%%%
%%%%%%%%%%%%%%%%%%%%%%%%%%%%%%%%       Method     %%%%%%%%%%%%%%%%%%%%%%%%%%%%%%%%%%%%
%%%%%%%%%%%%%%%%%%%%%%%%%%%%%%%%%%%%%%%%%%%%%%%%%%%%%%%%%%%%%%%%%%%%%%%%%%%%%%%%%%%%%%

\section{Method}
\label{sec:method}

\subsection{System Overview of Our Ranking Model}
\label{sec:method_components}
We use Progressive Layered Extraction (PLE) \cite{ref18} as the ranking model for multi-task prediction, 
including valid consumption task, watching time task and other tasks. Building on PLE, DUIA includes three components: 
User Interest Enhancement (UIE), User Positive Behavior Enhancement (UPBE) and User History Sequence Enhancement (UHSE), 
designed for enhancing user interest, as shown in Figure \ref{fig:f3}. 
A user's positive behavior indicates that the user is interested in the video, such as valid consumption, long watch time, interaction, and so on.

\subsection{User Interest Enhancement (UIE)}
\label{sec:upe}
User inrerest contains user profile and user history behavior sequences, as shown in Table \ref{table:user_profile} and Figure \ref{fig:f2}. 
To enhance user interest for both low-active and high-active users, the idea of UIE is to adopt an end-to-end dynamic stream clustering algorithm 
that groups similar users into clusters whose centers are stored for retrieval during prediction, building on existing MTL models.

To realize that, we investigate main clustering algorithms such as K-means and Divisive \cite{ref37}, 
as well as other algorithms like VQ-VAE \cite{ref38} and FSQ \cite{ref39}. 
However, VQ-VAE and FSQ are specifically designed for the requirements of CV, 
such as high-dimensional data, and involve processes like discretization, encoding, and decoding. 
As a result, they are complex and not fully aligned with the needs of UIE. 
Meanwhile, K-means and Divisive cannot be directly implemented in an end-to-end manner. 
To solve this problem, we propose an efficient Gradient-based Hierarchical Clustering Algorithm (GHCA) tailored for DUIA. 
GHCA performs hierarchical stream clustering using a gradient-based approach and adopts memory networks \cite{ref33} for the storage, updating, 
and retrieval of cluster centers, which is more efficient during end-to-end training, as shown in Figure \ref{fig:f3}. 
By utilizing hierarchical clustering, similar users can be grouped into multiple levels of granularity, 
allowing for generating enhancement vectors at different granularities and mitigating potential collapse problem that may occur during clustering.

Before streaming clustering, to obtain fixed-length vector of user interest, 
a simple auxiliary model is introduced, including a user tower and an item tower. The user tower takes user interest features as input, 
while the item tower uses item-related features as input. 
To save resource, the auxiliary model is shared by all the tasks of MTL model and its features' embedding is obtained from the MTL model. 
To better train this auxiliary model, the valid consumption label is used as supervision single, 
with the merge score that is generated by the predictions of all MTL tasks during online service as the weight for positive samples, 
while the weight for negative samples is equal and set to $1$, as shown in Eq. 1.
Moreover, to avoid affecting the main model's performance, the gradients from the auxiliary models are not back-propagated to the MTL model.

% cluster and update
In our RS, GHCA adopts a two-level hierarchical clustering structure: the first level is divided into $512$ clusters, 
and the second level includes $512 * 10$ clusters, whose centers are stored, updated and retrieve in the global user memory networks.
Before training, the centers of these clusters are randomly initialized. 
During training, the user interest vector $\upsilon_{user}$ is first clustered into the most similar cluster at the first level, 
and then further clustered into the most similar cluster within the corresponding second-level clusters, as shown in Eq. 2-5. 
In this paper, $\sigma$ is equivalent to sigmoid. 
When clustering into the corresponding clusters, a gradient-based clustering manner is used which is simple and efficient. 
In addition, to balance the influence of different users on clustering and reduce cost, 
we perform sampling based on user's active type and about 30\% of all samples participate in clustering. 

% prediction
When prediction, the most similar cluster center $\mu_{uie1, i}$ to the user interest vector $\upsilon_{user}$ is retrieved from the first-level clusters 
based on the similarity between $\upsilon_{user}$ and the cluster centers, 
and the most similar cluster center $\mu_{uie2, j}$ is also retrieved from the second-level clusters. 
If $\mu_{uie1, i}$ or $\mu_{uie2, j}$ is negatively correlated with $\upsilon_{user}$, the process of UIE will be stopped. 
Based on the similarity between $\mu_{uie1, i}$ with the user vector $\upsilon_{user}$, the weighted vector is calculated as shown in Eq. 6 
and then is concatenated with a personal vector to serve as the input to the generalization layer similar to PEPNet \cite{ref3}. 
The output is used as an additional enhancement vector for the main model. The same calculation approach is also applied to cluster center $\mu_{uie2,j}$. 
The personal vector is defined for each user to perform personalized fine-tuning of the most similar cluster center $\mu_{uie1, i}$ and $\mu_{uie2, j}$. 
\begin{align}
  \mathcal{L}_{auxiliary} &= -\frac{1}{n} \sum_{i=1}^{n} \left[\omega_{i} \ast y_i \ast \log(\hat{y}_i) + (1-y_i)\log(1-\hat{y}_i) \right] \,. \nonumber \\
  \omega_{i} &= (1.0 + sigmoid(merge\_score_{i})) \,. \\
  \hat{y}_i &= sigmoid(\upsilon_{user} \ast \upsilon_{item}) \,. \nonumber 
\end{align}
\begin{equation}
  i = \mathop{\arg\max}\limits_{i \in 1, ..., k_1} (\upsilon_{user} \ast [\mu_{uie1,1}, \mu_{uie1,2}, ..., \mu_{uie1,k_1}]^T) \,. 
\end{equation}
\begin{equation}
  j = \mathop{\arg\max}\limits_{j \in 1, ..., k_2} (\upsilon_{user} \ast [\mu_{uie2,1}, \mu_{uie2,2}, ..., \mu_{uie2,k_2}]^T) \,. 
\end{equation}
\begin{equation}
  \mathcal{L}_{\mathrm{uie\_l1}} = \frac{1}{n} \sum_{s=1}^{n} \sigma(\upsilon_{user} * \mu_{uie1, i}) \| \upsilon_{user} - \mu_{uie1, i} \|^2  \,.
\end{equation}
\begin{equation}
  \mathcal{L}_{\mathrm{uie\_l2}} = \frac{1}{n} \sum_{s=1}^{n} \sigma(\upsilon_{user} * \mu_{uie2, j}) \| \upsilon_{user} - \mu_{uie2, j} \|^2  \,.
\end{equation}
\begin{equation}
    \upsilon_{weighted} = sigmoid(\mu_{uie} \ast \upsilon_{user}) \ast \mu_{uie} \,.
\end{equation}

\subsection{User Positive Behavior Enhancement (UPBE)}
\label{sec:upbe}
To further enhance the user's interest, a personalized interest enhancement vector for each user is generated by UPBE 
from the perspective of user positive behaviors, as a supplement to UIE. 
The idea behind UPBE is that if a user exhibits positive behavior to a video, such as valid consumption, liking or sharing, 
the vector of the item, which is output by item tower of the auxiliary model, 
is used to search for the cluster centers that most likely to watch this video in the global user memory networks of UIE, 
and then the information contained in the corresponding cluster centers is extracted into the user's personalized interest enhancement vector 
via a gradient-based manner, as shown in Figure \ref{fig:f3}

% select approach, filter
The detailed process of UPBE is as follows: First, if the valid consumption label of a training sample is true, 
the vector output by the item tower of the auxiliary model is defined as $\upsilon_{item}$. Otherwise, the process is stopped. 
$\upsilon_{item}$ is then used to search for the user clusters most likely to watch the video within the global user memory networks in UIE, 
as shown in Eq. 7-8. 
After obtaining the two cluster centers $\mu_{uie1, i}$ and $\mu_{uie2, j}$ that are most likely to watch the video, 
it is also necessary to filter out cluster centers that are negatively correlated with the user vector $\upsilon_{user}$ based on their similarity. 
If they are negatively correlated, the process is stopped. 
Finally, the information included in  $\mu_{uie1, i}$ and $\mu_{uie2, j}$ are extracted into the user's personalized interest enhancement vectors 
through a gradient-based manner, as shown in Eq. 9-10.
During prediction, the enhancement vectors corresponding to the user is retrieved from the personalized user memory networks. 
After passing through the generalization layer, the outputs are used as a feature for the main model.
In this way, the user clusters most likely to watch the videos viewed by a user are used to enhance the user.
% add a Eq and explain; extract eq and its weight; filter eq. 
\begin{equation}
  i = \mathop{\arg\max}\limits_{i \in 1, ..., k_1} (\upsilon_{item} \ast [\mu_{uie1,1}, \mu_{uie1,2}, ..., \mu_{uie1,k_1}]^T) \,.
\end{equation}
\begin{equation}
  j = \mathop{\arg\max}\limits_{j \in 1, ..., k_2} (\upsilon_{item} \ast [\mu_{uie2,1}, \mu_{uie2,2}, ..., \mu_{uie2,k_2}]^T) \,.
\end{equation}
\begin{equation}
  \mathcal{L}_{\mathrm{upbe\_l1}} = \frac{1}{n} \sum_{s=1}^{n} \sigma(\mu_{uie1, i} \ast \mu_{upbe1, user}) \| \mu_{uie1, i} - \mu_{upbe1, user} \| ^2  \,.
\end{equation}
\begin{equation}
  \mathcal{L}_{\mathrm{upbe\_l2}} = \frac{1}{n} \sum_{s=1}^{n} \sigma(\mu_{uie2, j} \ast \mu_{upbe2, user}) \| \mu_{uie2, j} - \mu_{upbe2, user} \|^2  \,.
\end{equation}

\subsection{User History Sequence Enhancement (UHSE)}
\label{sec:uhse}
User history behavior sequences, including valid consumption sequences, interaction sequences, and others, 
play a crucial role in the performance of MTL models. 
Consequently, there has been extensive research focused on modeling user history behavior sequences. 
For simplicity, we take user valid consumption sequence as an example, as shown in Figure \ref{fig:f2}. 
As previously mentioned, there is a significant difference in the length of historical behavior sequences 
between low-activity users and high-activity users. 
If the amount of user history behaviors in the sequences is insufficient, the existing sequence modeling methods cannot perform effectively.
Therefore, UHSE is designed to enhance users' history sequence, especially for low-active users. 
The idea behind UHSE is to construct history behavior enhancement sequences for user history sequences, 
which are generated by searching for the most similar item cluster center in global item memory networks 
for each item in the user's history behavior sequences, as shown in Figure \ref{fig:f3}.
% cluster approach, using GHCA, add weight via similarty: eq explain
\begin{align}
  i = \mathop{\arg\max}\limits_{i \in 1, ..., k_3} (\upsilon_{item} \ast [\mu_{uhse1,1}, \mu_{uhse1,2}, ..., \mu_{uhse1,k_3}]^T) \,. \\
  j = \mathop{\arg\max}\limits_{j \in 1, ..., k_4} (\upsilon_{item} \ast [\mu_{uhse2,1}, \mu_{uhse2,2}, ..., \mu_{uhse2,k_4}]^T) \,. \\
  \mathcal{L}_{\mathrm{uhse\_l1}} = \frac{1}{n} \sum_{s=1}^{n} \sigma(\upsilon_{item} \ast \mu_{uhse1, i}) \| \upsilon_{item} - \mu_{uhse1, i} \| ^2  \,. \\
  \mathcal{L}_{\mathrm{uhse\_l2}} = \frac{1}{n} \sum_{s=1}^{n} \sigma(\upsilon_{item} \ast \mu_{uhse2, j}) \| \upsilon_{item} - \mu_{uhse2, j} \| ^2  \,.
\end{align}

Compared to directly clustering all the items in user history behavior sequences, which leads to excessive computation, 
we propose a novel and efficient approach. 
First, during training, samples with positive labels are selected, 
and the vector output by the item tower of the auxiliary model corresponding to each sample is defined as $\upsilon_{item}$. 
Then, during end-to-end training, these vectors are dynamically clustered in a streaming manner using the GHCA algorithm, as shown in Eq. 11-14. 
Each vector is clustered into the most similar cluster and all item clusters are organized into two layers, 
whose centers are stored, updated, and queried using global item memory networks.
When prediction, the user history enhancement sequences are used to calculate the user history enhancement vectors.

% over all loss and weight
In summary, the final loss of the ranking model is shown as in the Eq. 15, $\mathcal{L}_{MTL}$ is the loss of the MTL model, 
$\mathcal{L}_{auxiliary}$ is the loss of the auxiliary model, and $\rho$, $\lambda$ and $\eta$ 
are weight of clustering losses that should be set to small positive values to avoid fluctuation during clustering. In our RS, these values are set to 0.05.
\begin{align}
  \mathcal{L} &= \mathcal{L}_{MTL} + \mathcal{L}_{auxiliary} \nonumber \\
  &+ \rho \ast(\mathcal{L}_{\mathrm{uie\_l1}} + \mathcal{L}_{\mathrm{uie\_l2}})   \\
  &+ \lambda \ast (\mathcal{L}_{\mathrm{upbe\_l1}}  + \mathcal{L}_{\mathrm{upbe\_l2}}) \nonumber \\
  &+ \eta	\ast (\mathcal{L}_{\mathrm{uhse\_l1}} + \mathcal{L}_{\mathrm{uhse\_l2}}) \,. \nonumber
\end{align}

\subsection{Extensions for Items}
\label{sec:tail_items}
% highligh, limited to pages, for all items. add result in experiment.
The model performance on tail items is also poor due to highly skewed item distribution, which is usual in large-scale RSs.  
As mentioned before, in our RS, among all items exposed to users within a day, the tail 90\% items only account for less than 10\% of total exposures. 
To alleviate the problem of highly skewed item distribution, DUIA are also used for items, 
which also yield significant improvements. Due to page limitations, a detailed introduction is not provided in this paper.

\subsection{Extensions for Cold-Start}
\label{sec:cold_start}
% highligh, limited to pages, for all items. add result in experiment.
Cold start including new user start and new item start is also a very challenging problem in RS. 
DUIA and its variants have also been used to address cold-start problem in our RS, achieving outstanding results.
For example, UIE improves UVC by 7.57\% and UDT by 8.43\% on new users. 
In addition, UIE also effectively increases the distribution of new items. 
Due to space limitations, this paper does not discuss it in detail.

%%%%%%%%%%%%%%%%%%%%%%%%%%%%%%%%%%%%%%%%%%%%%%%%%%%%%%%%%%%%%%%%%%%%%%%%%%%%%%%%%%%%%%
%%%%%%%%%%%%%%%%%%%%%%%%%%%%%%       Experiments     %%%%%%%%%%%%%%%%%%%%%%%%%%%%%%%%%
%%%%%%%%%%%%%%%%%%%%%%%%%%%%%%%%%%%%%%%%%%%%%%%%%%%%%%%%%%%%%%%%%%%%%%%%%%%%%%%%%%%%%%
\section{Experiments}
\label{sec:experiments}

\subsection{Dataset and Evaluation Metrics}
\label{sec:dataset}
For offline experiments, we perform experiments on both public Taobao dataset and an industrial dataset collected from a large-scale RS. 

\textbf{Taobao Dataset}: Taobao dataset \cite{ref36} released by Alibaba is commonly used for offline experiments. 
Each training sample in this dataset consists of user ID, item ID, category ID, behavior type, and timestamp. 
According to the timestamp, it is splited into training set (80\%) and test set (20\%). 

\textbf{Industrial Dataset}: This dataset is collected from an industrial RS, serving hundreds of millions of users. 
The data collected during the first $14$ days is used for training including about 6.2 billion samples, 
and the data collected during the following $2$ days is used for testing. 
For simplicity, we take the valid consumption task of MTL models for comparison.

\textbf{Offline Evaluation Metrics}: For offline experiments, we adopt test Area Under Curve (AUC) to evaluate the performance of the compared models, 
which is widely used by related works. 

\textbf{Online Evaluation Metrics}: For online experiments, we deploy the compared models in the industrial RS for one week to conduct online A/B tests. 
We adopt User Valid Consumption (UVC) and User Duration Time (UDT) to evaluate each model, 
which are the two most important online metrics in our RS. In addition, we also compare diversity metrics of these models, 
including User Consumption on Category 1 (UCC1), User Consumption on Category 2 (UCC2) and User Consumption on Tag (UCT).

\begin{itemize}
  \item \textbf{UVC} is the average of all users' total valid consumptions during a day.
  \item \textbf{UDT} is the average of all users' total watching time within a day.
  \item \textbf{UCC1} is used to evaluate the distribution of category 1 of the videos watched by users.
  \item \textbf{UCC2} is used to evaluate the distribution of category 2 of the videos watched by users.
  \item \textbf{UCT} is used to evaluate the distribution of tags of the videos watched by users.
\end{itemize}

\subsection{Compared Methods}
\label{sec:compare}
We compare DUIA with two typical methods \cite{ref28, ref29} proposed by Google. 
Since the methods such as \cite{ref3, ref31} are orthogonal to DUIA, there is no need to compare them with DUIA. 
Furthermore, we design multiple variants of DUIA to illustrate the effects of each enhancement component on performance respectively. 
For a fair comparison, all compared methods use the same features as input and adopt the same optimizer Adam. 
All the compared methods are introduced as follows:
\begin{itemize}
  \item \textbf{PLE} \cite{ref18}: PLE is proposed by Tencent which exhibits excellent performance. It is used as the baseline for comparison.

  \item \textbf{PLE with CL}: A CL enhancement component is developed for low-active users according to \cite{ref28} building on PLE.
  
  \item \textbf{PLE with TL}: A TL enhancement component is developed for low-active users according to \cite{ref29} building on PLE.
  
  \item \textbf{UIE}: UIE is implemented based on PLE, which is used for enhancing user interest for all users.
  
  \item \textbf{UIE and UPBE}: UPBE is developed based on PLE with UIE.
  
  \item \textbf{DUIA}: DUIA including UIE, UPBE and UHSE is implemented based on PLE, which is designed to validate 
  the performance of all enhancement components for all users.
  
  \item \textbf{DUIA for Users and Items}: DUIA is used for both users and items, aiming to further improve performance.
\end{itemize}

The parameter settings of DUIA are shown in Table \ref{tab:param_set}. 
The abbreviations are: NFC for the Number of First-level Clusters, NSC for the Number of Second-level Clusters, and dims for dimensions.
\begin{table}[hbtp!]
    \caption{The parameters setting of DUIA.}
    \label{tab:param_set}
    \renewcommand\arraystretch{1.0}{
    \begin{minipage}{\columnwidth}
    \begin{center}
    \resizebox{\linewidth}{!}{
    \begin{tabular}{c|c}
    \toprule
    \textbf{\qquad \qquad \qquad Parameter \qquad \qquad \qquad} & \textbf{\qquad \qquad Value \qquad \qquad}  \\ 
    \hline
    \textbf{the towers of the auxiliary model}  & 256 * 128 * 64   \\
    \textbf{the NFC in UIE} & 512  \\
    \textbf{the NSC in UIE} & 10  \\
    \textbf{the NFC in UHSE} & 2048 \\
    \textbf{the NSC in UHSE} & 10 \\
    \textbf{the dims of all memory networks} & 64 \\ 
    \textbf{$\rho$} & 0.05 \\ 
    \textbf{$\lambda$} & 0.05 \\ 
    \textbf{$\eta$} & 0.05 \\ 
    \bottomrule
    \end{tabular}
    }
    \end{center}
    \end{minipage}
    }
    \vspace{-0.15in}
\end{table}

\subsection{Offline Evaluation}
\label{sec:offline}
% update, the setting of the model
In this section, extensive offline experiments are conducted to demonstrate the remarkable performance of DUIA compared to other methods. 
All the models are tested on the same dataset and the results are shown in Table \ref{tab:taobao_auc}.

% public dataset
\subsubsection{The Results on Taobao Dataset}
\label{sec:taobao_result}
\begin{table}[hbtp!]
    \caption{The test AUC of the models on Taobao dataset.}
    \label{tab:taobao_auc}
    \renewcommand\arraystretch{1.0}{
    \begin{minipage}{\columnwidth}
    \begin{center}
    \resizebox{\linewidth}{!}{
    \begin{tabular}{c|c}
    \toprule
    \textbf{\qquad \qquad \qquad Models \qquad \qquad \qquad} & \textbf{\qquad \qquad Test AUC \qquad \qquad}  \\ 
    \hline
    \textbf{PLE}  & 0.8836    \\
    \textbf{UIE} & 0.8940  \\
    \textbf{UIE and UCBE} & 0.8968  \\
    \textbf{DUIA} & \textbf{0.9003} \\
    \textbf{DUIA for Users and Items} & \textbf{0.9041} \\
    \bottomrule
    \end{tabular}
    }
    \end{center}
    \end{minipage}
    }
\end{table}

Since Taobao dataset does not include features about user types (low-active or high-active), 
we evaluate the models' performance for all users. 
In addition, PLE with CL and PLE with TL designed for low-active users are not evaluated on this dataset.
The results of all the models on Taobao dataset are shown in Table \ref{tab:taobao_auc}. Each experiment is repeated $3$ times. 

Compared with the baseline, UIE significantly improves model performance, with an increase of 1.04\%. 
By constructing personalized enhancement vectors for users based on positive behaviors, UPBE further improves model performance, 
In addition, UHSE enhances user history sequences using the most similar item centers stored in the global item memory networks. 
DUIA including all the above components achieves remarkable improvement, achieving an increase of 1.67\% over the baseline. 
Moreover, DUIA for users and items achieves the largest improvement, with a 2.05\% increase over the baseline.

% industrial dataset
\subsubsection{The Results on Industrial Dataset}
\label{sec:offline_industrial_result}
\begin{table}[hbtp!]
    \caption{The test AUC of the models on the industrial dataset. 
    {\textquotesingle}-{\textquotesingle} represents there are no modification compared to baseline.}
    \label{table:industrial_auc}
    \renewcommand\arraystretch{1.1}{
    \begin{minipage}{\columnwidth}
    \begin{center}
    \resizebox{\linewidth}{!}{
    \begin{tabular}{ c|c|c }
    \toprule
    \textbf{Models} & \textbf{Low-active Users} & \textbf{High-active Users} \\ \hline
    \textbf{PLE}  & 0.7842  & 0.8022   \\
    \textbf{PLE with CL} & 0.7846 & - \\
    \textbf{PLE with TL} & 0.7849 & - \\
    \textbf{UIE} & 0.7886 &  0.8047 \\
    \textbf{UIE and UCBE} & 0.7918 &  0.8073 \\
    \textbf{DUIA} & \textbf{0.7961}  & \textbf{0.8119} \\
    \textbf{DUIA for Users and Items} & \textbf{0.8014}  & \textbf{0.8153} \\
    \bottomrule
    \end{tabular}
    }
    \end{center}
    \end{minipage}
    }
  \end{table}

% online results
\begin{table*}[hbtp!]
  \caption{The results of online A/B experiments. {\textquotesingle}-{\textquotesingle} represents that there are no modifications compared to baseline.}
  \label{tab:online_results}
  \small
  \renewcommand\arraystretch{1.05}{
  \centering
  % \begin{tabular}{c|cccc|cccc}
  \resizebox{\textwidth}{!}{
  \begin{tabularx}{\textwidth}{c|YYYYY|YYYYY}
  % \begin{tabularx}{\textwidth}{c|ccccc|ccccc}
  \toprule
    &
    \multicolumn{5}{c|}{\textbf{Low-active Users}} &
    \multicolumn{5}{c}{\textbf{High-active Users}} \\ \cline{2-11}
    \multirow{-2}{*}{\textbf{Compared Models} } &
    \multicolumn{1}{c}{ \textbf{UVC} } &
    \multicolumn{1}{c}{ \textbf{UDT} } &
    \multicolumn{1}{c}{ \textbf{UCC1} } &
    \multicolumn{1}{c}{ \textbf{UCC2} } &
    \multicolumn{1}{c|}{ \textbf{UCT} } &
    \multicolumn{1}{c}{ \textbf{UVC} } &
    \multicolumn{1}{c}{ \textbf{UDT} } &
    \multicolumn{1}{c}{ \textbf{UCC1} } &
    \multicolumn{1}{c}{ \textbf{UCC2} } &
    \multicolumn{1}{c}{ \textbf{UCT}  } \\ 
    \hline
    \textbf{PLE} &
    $\ast$ &
    $\ast$ &
    $\ast$ &
    $\ast$ &
    $\ast$ &
    $\ast$ &
    $\ast$ &
    $\ast$ &
    $\ast$ &
    $\ast$
    \\

    \textbf{PLE with CL} &
    +0.51\% &
    +0.39\% &
    +0.07\% &
    +0.68\% &
    +0.79\% &
    - &
    - &
    - &
    - &
    -
    \\

    \textbf{PLE with TL} &
    +0.80\% &
    +0.43\% &
    +0.13\% &
    +0.92\% &
    +0.86\% &
    - &
    - &
    - &
    - &
    -
    \\

    \textbf{UIE} &
    +3.49\% &
    +2.01\% &
    +1.15\% &
    +1.97\% &
    +2.13\% &
    +2.19\% &
    +1.46\% &
    +0.64\% &
    +1.42\% &
    +1.83\%
    \\

    \textbf{UIE and UCBE} &
    +6.03\% &
    +2.94\% &
    +3.23\% &
    +3.25\% &
    +3.96\% &
    +3.26\% &
    +2.29\% &
    +1.25\% &
    +2.13\% &
    +2.97\%
    \\

    \textbf{DUIA} &
    +8.12\% &
    +4.05\% &
    +3.96\% &
    +4.60\% &
    +5.23\% &
    +4.33\% &
    +3.01\% &
    +1.71\% &
    +3.29\% &
    +3.74\%
    \\

    \textbf{DUIA for Users and Items} &
    +9.68\% &
    +5.74\% &
    +3.98\% &
    +4.64\% &
    +5.42\% &
    +5.60\% &
    +3.88\% &
    +1.76\% &
    +3.38\% &
    +4.02\%
    \\

  \bottomrule

  \end{tabularx}
  }}
  %\vspace{-2mm}
\end{table*} 

The results of the models on the industrial dataset are shown in Table \ref{table:industrial_auc}.
The AUC of PLE with CL on low-active users is slightly higher than that of PLE with an improvement of 0.04\%.
Due to CL \cite{ref28} easily generating meaningless data in the context of RS, the augmentation data has limit value to the model. 
The AUC of PLE with TL on low-active users is 0.07\% higher than that of the baseline. 
UIE significantly improves model performance for low-active users, with an improvement of 0.44\%. 
Moreover, it also enhances model performance for high-active users, with an improvement of 0.25\%. 
UPBE and UHSE further improve performance by constructing personalized enhancement vectors and history enhancement sequences for users, respectively. 
DUIA, which incorporates UIE, UPBE, and UHSE, achieves remarkable improvements, increasing AUC by 1.19\% on low-active users and 
0.97\% on high-active users compared to the baseline model. 
DUIA for users and items achieves the greatest improvements, 
with increases of 1.72\% on low-active users and 1.31\% on high-active users over the baseline model.
Other tasks also show similar improvements.

\subsubsection{Hyper-Parameter Analysis}
\label{sec:hyper_analysis}
The hyper-parameters of DUIA need to be carefully selected according to specific recommendation scenario.
We conduct an analysis of the relationship between the number of first-level clusters and performance in UIE and UHSE through experiments, 
using the industrial dataset. 
For simplicity, the number of second-level clusters is set to be ten times that of the first-level clusters. 
The results are shown in the Table \ref{table:cluster_num}. As the number of cluster centers increases, 
the model performance of UIE and UHSE also improves; however, after reaching a certain number, the performance gains become slow.
\begin{table}[hbtp!]
    \caption{The relationship between the Number of First-level Clusters (NFC) and the performance in UIE and UHSE.}
    \label{table:cluster_num}
    \small
    \renewcommand\arraystretch{1.0}{
    \begin{minipage}{\columnwidth}
    \begin{center}
    \resizebox{\linewidth}{!}{
    \begin{tabular}{c|c|c|c}
        \toprule
        \quad \textbf{Method} \quad & \quad \textbf{NFC} \quad \quad & \textbf{Low-active Users} & \textbf{High-active Users} \\ \hline
        \multirow{5}{*}{\textbf{UIE}} & 128 & 0.7859  & 0.8033 \\
                             & 256 & 0.7875  & 0.8041 \\
                             & 512 & 0.7886 & 0.8047 \\
                             & 1024 & 0.7890 & 0.8049 \\
                             & 2048 & 0.7893 & 0.8050 \\
        \hline 
        \multirow{5}{*}{\textbf{UHSE}} & 512 & 0.7872 & 0.8055 \\
                             & 1024 & 0.7881 & 0.8063 \\
                             & 2048 & 0.7889 & 0.8071 \\
                             & 4096 & 0.7891 & 0.8076 \\
                             & 8192 & 0.7893 & 0.8078 \\
        \bottomrule
    \end{tabular}
    }
    \end{center}
    \end{minipage}
    }
\end{table}

We also analyze the relationship between the dimensions of the memory networks and model performance in DUIA. 
It can be seen that as the dimension increases from 24, the model performance also improves; 
however, after the dimension exceeds 128, the improvement becomes gradual, as shown in Table \ref{table:dim_num}. 
The analysis of other parameters will not be listed one by one due to page limit.
\begin{table}[hbtp!]
  \caption{The relationship between the Dimensions of the Memory Networks (DMN) and the performance in DUIA.}
  \label{table:dim_num}
  \small
  \renewcommand\arraystretch{1.0}{
  \begin{minipage}{\columnwidth}
  \begin{center}
  \resizebox{\linewidth}{!}{
  \begin{tabular}{c|c|c|c}
      \toprule
      \quad \textbf{Method} \quad & \quad \textbf{DMN} \quad \quad & \textbf{Low-active Users} & \textbf{High-active Users} \\ \hline
      \multirow{5}{*}{\textbf{DUIA}} & 24 & 0.7948  & 0.8109 \\
                           & 48 & 0.7956  & 0.8116 \\
                           & 64 & 0.7961 & 0.8119 \\
                           & 128 & 0.7963 & 0.8120 \\
                           & 256 & 0.7963 & 0.8121 \\
      \bottomrule
  \end{tabular}
  }
  \end{center}
  \end{minipage}
  }
  \vspace{-0.1in}
\end{table}

\subsubsection{cluster Analysis}
\label{sec:cluster_analysis}
To verify the effectiveness of clustering, we analyze the hit rate under different numbers of clusters. 
Taking UIE as an example, we first train UIE on the industrial dataset and 
then calculate the proportion of the first-level clusters that are hit during testing. Each experiment is repeated 5 times. 
As shown in Table \ref{table:hit_rate}, as the number of clusters in the first layer increases, 
the hit rate during testing decreases, which is widespread in clustering algorithms. 
To alleviate this problem, the GHCA algorithm proposed in this work adopts a multi-level clustering approach. 
\begin{table}[hbtp!]
  \caption{The hit rate of the first-level clusters in UIE.}
  \label{table:hit_rate}
  \small
  \renewcommand\arraystretch{1.0}{
  \begin{minipage}{\columnwidth}
  \begin{center}
  \resizebox{\linewidth}{!}{
  \begin{tabular}{c|c|c}
      \toprule
      \qquad \textbf{Method} \qquad \quad & \qquad \textbf{NFC in UIE} \qquad \quad & \qquad \qquad \textbf{Hit Rate} \qquad \qquad \quad \\ \hline
      \multirow{5}{*}{\textbf{UIE}} & 128 & 0.93  \\
                           & 256 & 0.85  \\
                           & 512 & 0.79  \\
                           & 1024 & 0.73 \\
                           & 2048 & 0.68  \\
      \bottomrule
  \end{tabular}
  }
  \end{center}
  \end{minipage}
  }
\end{table}

\subsection{Online Evaluation}
\label{sec:online}
The online results are presented in Table \ref{tab:online_results}, and all improvements are statistically significant with $p$-values less than 0.05.
PLE with CL and PLE with TL slightly improve UVC and UDT compared to the baseline. UIE increases UVC by 3.49\% and UDT by 2.01\% 
for low-active users, and UVC by 2.19\% and UDT by 1.46\% for high-active users, achieving significant improvements over the baseline. 
Moreover, UIE also improves diversity metrics for all users, especially for low-active users, indicating that the user's interest are effectively enhanced. 
UIE and UCBE further improve the online results compared to UIE alone.
DUIA achieves remarkable improvements, with increases of 8.12\% in UVC and 4.05\% in UDT for low-active users, and 4.33\% in UVC and 
3.01\% in UDT for high-active users. DUIA applied to both users and items achieves the best results, with improvements of 9.68\% in UVC and 
5.74\% in UDT for low-active users, and 5.60\% in UVC and 3.88\% in UDT for high-active users. 
Furthermore, the diversity metrics are also significantly improved.

%%%%%%%%%%%%%%%%%%%%%%%%%%%%%%%%%%%%%%%%%%%%%%%%%%%%%%%%%%%%%%%%%%%%%%%%%%%%%%%%%%%%%%
%%%%%%%%%%%%%%%%%%%%%%%%%%%%%%       Conclusion     %%%%%%%%%%%%%%%%%%%%%%%%%%%%%%%%%%
%%%%%%%%%%%%%%%%%%%%%%%%%%%%%%%%%%%%%%%%%%%%%%%%%%%%%%%%%%%%%%%%%%%%%%%%%%%%%%%%%%%%%%
\section{Conclusion}
\label{conclusion}
In this work, we first highlight the problem that lots of users in industrial RSs merely have sparse interest, 
resulting in poor recommendation performance for them, which has long been overlooked. We propose DUIA to solve this challenging problem.
The key idea of DUIA is to construct enhancement vectors for enhancing user interest 
from different perspectives, using the most similar centers and the most relevant centers which are generated by dynamic 
stream clustering in memory networks. 
In addition, we also design GHCA for DUIA, which performs end-to-end clustering via gradient descent efficiently. 
By generating enhancement information for user interest, DUIA remarkably improves the performance of ranking model on low-active users and 
achieves significant improvements on high-active users as well. 
Moreover, DUIA is also used for long-tail items and cold-start problem. 
Until now, DUIA has been deployed in multiple large-scale RSs and has also achieved significant improvements.

\section*{GenAI Usage Disclosure}
I know that the ACM's Authorship Policy requires full disclosure of all use of generative AI tools in all stages of the research 
(including the code and data) and the writing. No GenAI tools were used in any stage of the research, nor in the writing.

% \bibliographystyle{ACM-Reference-Format}
% \balance
% \bibliography{main}

\end{document}